\documentclass[aps,prd,twocolumn,floatfix, preprintnumbers,nofootinbib]{revtex4}
\usepackage[pdftex]{graphicx}
\usepackage[mathscr]{eucal}
\usepackage[pdftex]{hyperref}
\usepackage{amsmath,amssymb,bm,yfonts}
\usepackage[utf8]{inputenc}
\usepackage{xcolor}

\newif\ifarxiv
\arxivfalse

\begin		{document}

\def\del	{\nabla}

\def\K {\mathcal K}

\def\g {\frak g}
\def\h {\frak h}
\def\H {\frak H}
\def\E {E}

\def\suck[#1]#2{\includegraphics[#1]{#2}}        

\title
    {
Holographic duality and mode stability of de Sitter space in semiclassical gravity 
    }

\author{Paul~M.~Chesler}
\email{pchesler@g.harvard.edu}

\author{Abraham~Loeb}
\email{aloeb@cfa.harvard.edu}

\affiliation
    {Black Hole Initiative, Harvard University, Cambridge, MA 02138, USA}

\date{\today}

\begin{abstract}
	We employ holographic duality to compute $\langle T_{\mu \nu} \rangle$ in strongly coupled $\mathcal N = 4$ supersymmetric Yang-Mills theory and then study evolution of the semiclassical Einstein field equations sourced by $\langle T_{\mu \nu} \rangle$.  Linearizing about de Sitter space, we find that the semiclassical equations of motion reduce to a four dimensional scalar wave equation coupled to a five dimensional scalar wave equation.  We compute the mode spectrum of these equations and find that there exists a critical value of the Hubble constant $H_c$ for which de Sitter space is unstable when $H < H_c$ and mode stable when $H > H_c$.  
\end{abstract}


\maketitle
\parskip	2pt plus 1pt minus 1pt

\section{Introduction}

General Relativity (GR) and quantum field theory  are two pillars of modern physics, with GR describing the large scale structure of spacetime and quantum field theories describing the microscopic properties of matter.
Combining these two very different descriptions of nature into a single quantum theory of gravity remains an unsolved problem.  A less ambitions but still interesting problem is the interaction of matter quantum fields and gravity at scales much larger than the Planck length.  

An analogous question can be posed in electrodynamics:  how do matter quantum fields interact with macroscopic electromagnetic fields? This question is addressed by the theory of Macroscopic Electrodynamics. The macroscopic equations of motion 
can be derived via phenomenological considerations or alternatively, at least for simple systems, via first principles calculations (see e.g. \cite{kapusta_gale_2006,fetter2012quantum}).  They read 
\begin{equation}
\label{eq:macroem}
\frac{1}{e^2} \del_\nu F^{\mu \nu} =   \langle J^\mu \rangle,
\end{equation}
where  $\langle J^\mu \rangle$ is the expectation value of the electric current, which is  a functional of the macroscopic vector potential $A_\mu$, and $e$ is the electric charge.
%
A reasonable expectation is that matter quantum fields couple to gravity in an analogous fashion, with macroscopic evolution governed by the semiclassical Einstein equations (see e.g. \cite{birrell_davies_1982})
\begin{equation}
\label{eq:semiclassicalEinstein}
R_{\mu \nu} - {\textstyle \frac{1}{2}} (R- 2 \Lambda) g_{\mu \nu} = 8 \pi G\langle T_{\mu \nu} \rangle,
\end{equation}
with $\Lambda$ being the cosmological constant, $G$ Newton's constant, and $\langle T_{\mu \nu} \rangle$ is the expectation value of the stress tensor, which is a functional of the metric $g_{\mu \nu}$.

In this paper we study the stability of the Poincare patch of $3+1$ dimensional de Sitter space, which is the maximally symmetric solution to the semiclassical equations of motion (\ref{eq:semiclassicalEinstein}).  It was long ago shown that de Sitter space is stable in classical GR \cite{friedrich1986,FRIEDRICH1986101,Anderson:2004ir}.  Subsequently it was shown that Minkowski space is also stable in classical GR \cite{Christodoulou:1993uv,Lindblad:2004ue}.  The same, however, does not hold in semiclassical gravity.  Numerous studies have found that de Sitter space and Minkowski space 
are unstable to exponentially growing modes
\cite{Horowitz:1978fq,Horowitz:1980fj,Hartle:1981zt,Suen:1989bg,Suen:1988uf,Jordan:1987wd,RandjbarDaemi:1981wd,Matsui:2018iez,Matsui:2019tlf,Matsui:2019tah}.%
\footnote{A notable exception is Ref.~\cite{Anderson:2002fk}, where it was argued that Minkowski space can be mode stable in semiclassical gravity provided the stress spectral density vanishes sufficiently rapidly in the IR.} Notably, instabilities can occur at arbitrarily long wavelength.

Thus far, stability analyses in semiclassical gravity have been limited to free or weakly interacting quantum field theories.  Here we shall study the opposite limit, employing a strongly interacting quantum field theory.  Indeed, it has been 
argued that interactions can destabilize de Sitter space (see e.g. Ref.~\cite{Polyakov:2012uc}).  Currently there exists only one theoretical framework capable of systematically studying dynamics in strongly coupled quantum field theories: holographic duality \cite{Maldacena:1997re}. 

Holographic duality maps the dynamics of certain strongly coupled non-Abelian gauge theories onto the dynamics of gravity in higher dimensions. The dual gravitational dynamics becomes classical in the limit of large number of colors $N$ in the gauge theory. Numerous authors have employed holographic duality to study quantum field theory in de Sitter space \cite{Maldacena:2012xp,Fischler:2013fba,Fischler:2014ama,Fischler:2014tka,Nguyen:2017ggc}, including semiclassical gravity \cite{Hawking:2000bb}.  Likewise, it has widely been employed to study the dynamics of strongly coupled quark-gluon plasma produced in heavy ion collisions at RHIC and the LHC (for a review see Ref.~\cite{CasalderreySolana:2011us}) and strongly coupled condensed matter systems (for a review see Ref.~\cite{Hartnoll:2016apf}).
The utility of the duality is that $\langle T_{\mu \nu} \rangle$ and other observables can be computed by solving classical partial differential equations, albeit in higher dimensions.  In fact, with holography the semiclassical equations of motion (\ref{eq:semiclassicalEinstein}) reduce to Einstein's equations in four dimensions coupled to Einstein's equations in five dimensions via boundary conditions. 
The simplest holographic quantum field theory is $\mathcal N = 4$ supersymmetric Yang-Mills theory (SYM), which is the theory we choose to study.  SYM is a conformal field theory and dual to gravity in five dimensional asymptotically anti-de Sitter space (AdS$_5$) \cite{Maldacena:1997re}.

Aside from being necessary to employ holography, the large $N$ limit also puts the semiclassical equations of motion (\ref{eq:semiclassicalEinstein}) on stronger theoretical footing \cite{Tomboulis:1977jk,Hartle:1981zt}. In particular, the large $N$ limit provides a clean separation between macroscopic scales where instabilities may occur, and microscopic scales where the semiclassical equations of motion break down.  Due  to the fact  that 
$\langle T_{\mu \nu} \rangle \sim N^2$, the semiclassical equations of motion (\ref{eq:semiclassicalEinstein}) have a well-defined $N \to \infty$ limit provided 
\begin{equation}
\label{eq:Gscaling}
G \sim \frac{1}{N^2}.
\end{equation}  
This scaling means the Planck length is order $1/N$.  However, the semiclassical equations of motion
should break down well before the Planck scale.  Just as macroscopic electrodynamics shouldn't apply at atomic scales, where fluctuations in the current become large, the semiclassical Einstein equations (\ref{eq:semiclassicalEinstein}) should not apply at scales where fluctuations in the stress tensor become large \cite{Kuo:1993if,Anderson:2002fk}.  In large $N$ SYM, connected stress correlators scale like $N^2/\Delta x^8$ where $\Delta x$ is the point separation. This means that fluctuations in the stress become large at length scales on the order of $N^{-1/4}$, which can be made arbitrarily small by taking $N \to \infty$. In contrast, with the scaling (\ref{eq:Gscaling}), the semiclassical equation of motion (\ref{eq:semiclassicalEinstein}) are independent of $N$, meaning that in the $N \to \infty$ limit, any finite wavelength instability lies within the domain of validity of the semiclassical equations of motion.

We linearize the semiclassical equations of motion about de Sitter space and study the spectrum of allowed modes, including inhomogenous and anisotropic modes. We find the linearized equations of motion reduce to a four dimensional scalar wave equation coupled to a five dimensional scalar wave equation.  The mode spectrum of these equations can be determined analytically.  We find that there exists a critical value of the Hubble constant $H_c$ for which de Sitter space is unstable when $H < H_c$ and mode stable when $H > H_c$.  The instablities result in both inhomogeneities and anisotropies growing exponentially fast.
Up to logarithmic corrections, the critical Hubble constant scales like $H_c^2 \sim \frac{1}{G N^2}$.
Hence, in the large $N$ limit $H_c^2$ is parametrically smaller than the Planck mass (and by the scaling relation (\ref{eq:Gscaling}), independent of $N$). 

We also study the mode stability of the semiclassical Maxwell equations (\ref{eq:macroem}) with strongly coupled quantum fields in de Sitter space.  Like our gravitational analysis, we find that the semiclassical Maxwell equations
are mode stable only if $H$ is greater than some critical value.  

An outline of the remainder of our paper is as follows.  In Sec.~\ref{sec:bdinvariants} we construct a diffeomorphism invariant 
formulation of the linearized semiclassical equations of motion (\ref{eq:semiclassicalEinstein}).  In Sec.~\ref{eq:holographic} we employ holographic duality to compute $\langle T_{\mu \nu} \rangle$ and derive the linearized semiclassical equations of motion.  In Sec.~\ref{eq:modesolutions} we study the spectrum of allowed modes and in Sec.~\ref{sec:dis} we discuss our results, including possible origins of and resolutions to  instabilities.  In the Appendix, we present our study of the mode stability of the semiclassical Maxwell equations.

\section{Gauge invariants}
\label{sec:bdinvariants}

In the Poincare patch the perturbed de Sitter metric may be written as
\begin{align}
\label{eq:perturbeddS}
ds^2 = - [1 {-} h_{00}]dt^2 + 2 h_{0i} dt dx^i +e^{2 H t} [ \delta_{ij} + h_{ij} ] dx^i dx^j,
\end{align}
where $h_{\mu \nu}$ are the metric perturbations.  The linearized semiclassical Einstein equations then take the schematic form
\begin{equation}
\label{eq:lineareinstein}
\Delta^{\alpha \beta}_{\ \ \mu \nu} h_{\alpha \beta} = 8 \pi G \langle \delta T_{\mu \nu} \rangle,
\end{equation}
where $\Delta^{\alpha \beta}_{\ \ \mu \nu}$ is a second order differential operator
and $\langle \delta T_{\mu \nu} \rangle$ is the change in the stress tensor induced by the metric perturbations.

Instead of working directly with $h_{\mu \nu}$, we have found 
it convenient to work with gauge invariant combinations of $h_{\mu \nu}$ and its derivatives.  This is analogous to working with electric and 
magnetic fields in electrodynamics instead of the gauge field.  The relevant gauge transforms are infinitesimal diffeomorphisms,
\begin{equation}
\label{eq:difftrans0}
x^\mu \to x^\mu + \xi^\mu,
\end{equation}
where $\xi^\mu$ is an arbitrary function of $x^\mu$.  Under this transformation the metric transforms like
\begin{equation}
\label{eq:difftrans}
g_{\mu \nu} \to g_{\mu \nu} - \del_{\mu} \xi_\nu -  \del_{\nu} \xi_\mu,
\end{equation}
where $\del_\mu$ is the covariant derivative under the background geometry.

Following Refs.~\cite{Kovtun:2005ev,Chesler:2007sv,Chesler:2007an,Hong:2011bd}, to construct gauge invariants we introduce a spatial Fourier transform and work with mode amplitudes $ h_{\mu \nu}(t,\bm q)$ with momentum vector $\bm q$.  Let $\{\hat {\bm q},\hat { \bm \epsilon}_1,\hat {\bm \epsilon}_2\}$ be an orthonormal basis of polarization vectors.  
We then decompose $h_{\mu \nu}$ in terms of longitudinal and transverse components as follows
\begin{subequations}
	\label{eq:polarizationframe}
\begin{eqnarray}
 h_{0i} &=&  h_{0q} \hat q^i +  h_{0a} \hat \epsilon_a^i,
\\
 h_{ij} &=&  h_{a q}\hat \epsilon_a^{(i} \hat q^{j)} +  h_{a b} \hat \epsilon_a^{i}\hat \epsilon_b^{j}.
\end{eqnarray}
\end{subequations}
Here and in what follows lower case latin indices $a,b,c$ from the beginning of the alphabet run over the two directions transverse to $\bm q$
whereas $i,j$ run over all the three spatial directions.
We also employ an identical polarization frame decomposition for the stress tensor perturbation $\langle \delta T_{\mu \nu} \rangle$, the 5D metric perturbation studied in Sec.~\ref{eq:5dpert}, and gauge fields studied in the Appendix.
Defining $\dot h_{\mu \nu} \equiv \partial_t h_{\mu \nu}$, it is easy to see that the following 
quantities are invariant under the transformation (\ref{eq:difftrans}),
\begin{subequations}
	\label{eq:gaugeinvariants}
	\begin{align}
     \nonumber
     W_0 \equiv &\, \ddot h_{aa} +  \ddot h_{qq} + 4 H [\dot h_{aa} + \dot h_{qq}]+  [12 H^2 {-} e^{-2 H t} q^2 ] h_{00}   \ \ \ \ \ \
     \\ 
     + &\,  3 H \dot h_{00}+e^{-2 H t} [q^2 h_{aa} {-} 4  i q h_{0q} {-} 2  i q \dot h_{0q}], 
	 \\
	 Z_0 \equiv &\, \textstyle 2 q^2 e^{-2 H t} h_{00} + 4 i q \, e^{-2 H t}[ \dot h_{0q}- H h_{0q}] 
	\\ \nonumber
	+ & \,\ddot h_{aa}- 2 \ddot h_{qq}+ H \dot h_{aa} - 2 H \dot h_{qq}+ e^{-2 H t} q^2 h_{aa},
	\\
	 Z_1 \equiv &\, [i q \dot h_{aq}+ q^2 e^{-2 H t} h_{0a}] \hat  {\bm \epsilon}_a ,
	\\
	 Z_2  \equiv &\, [\textstyle h_{ab} - \frac{1}{2} h_{cc} \delta_{ab}] \hat  {\bm \epsilon}_a \otimes \hat {\bm \epsilon}_b.
	\end{align}
\end{subequations}

The invariants $W_0$ and $Z_0$ transform as scalars under rotations about the $\bm q$ axis
whereas $Z_1$ and $Z_2$ transform as a vector and tensor, respectively.  
Altogether Eqs.~(\ref{eq:gaugeinvariants}) contain 6 gauge invariant degrees of freedom.
With $W_0$ and $Z_s$ known, and a gauge choice specified, the metric perturbations $h_{\mu \nu}$ can be reconstructed.  In particular, in the gauge $h_{0 \mu} = 0$, finiteness of the above gauge invariants guarantees finiteness of the metric perturbations themselves.  Therefore, demonstrating linear stability of de Sitter is tantamount to showing that the gauge invariants $W_0$ and $Z_s$ remain bounded as $t \to \infty$.

The equations of motion for the invariants follow from the linearized semiclassical equations of motion.
Consider first the trace of Eq.~(\ref{eq:semiclassicalEinstein}), which yields
\begin{equation}
\label{eq:traceeq}
-R + 4 \Lambda = 8 \pi G \langle T^\mu_{\ \mu} \rangle.
\end{equation}
The trace of the stress tensor is independent of the quantum state and fixed by the Weyl anomaly to be \cite{Duff:1993wm,Henningson:1998ey}
\begin{equation}
\label{eq:trace}
\langle T^{\mu}_{\ \mu} \rangle = \frac{N^2}{32 \pi^2} \left (R^{\mu \nu} R_{\mu \nu} - \frac{1}{3} R^2 \right).
\end{equation}
At zeroth order, Eqs.~(\ref{eq:traceeq}) and (\ref{eq:trace}) imply
\begin{equation}
-3 H^2 + \Lambda = - \frac{3  GN^2 }{4 \pi} H^4,
\end{equation}
or equivalently,
\begin{equation}
\label{eq:Hequation}
H^2 = \frac{2 \pi}{G N^2} \left [ 1 \pm \sqrt{1 - \frac{\Lambda GN^2}{3 \pi}} \right].
\end{equation}
By adjusting the cosmological constant, Eq.~(\ref{eq:Hequation}) implies one can get any value of $H^2 \leq H_{\rm max}^2$.  The maximum Hubble constant is simply,
\begin{equation}
\label{eq:Hmax}
	H_{\rm max} = \sqrt{\frac{4 \pi}{G N^2}}.
\end{equation}
At first order in $h_{\mu \nu}$, Eqs.~(\ref{eq:traceeq}) and (\ref{eq:trace}) imply
\begin{equation}
W_0 = 0.
\end{equation}
Hence $W_0$ is non-dynamical.

The equation of motion for $Z_s$ can be obtained by taking appropriate linear combinations of the linearized semiclassical equations (\ref{eq:lineareinstein}) and read
\begin{subequations}
	\label{eq:boundaryinvariants}
\begin{align}
\label{eq:Z0eq}
q^2 Z_0 = & \,  16 \pi G  \Big [   2 q^2 \langle \delta T_{00} \rangle - 6 H i q  \langle \delta T_{0q} \rangle  \\  \nonumber
 & \ \ \ \ \ \ \ +  \, q^2 e^{-2 H t} \left ( \langle \delta T_{aa} \rangle - 2 \langle \delta T_{qq} \rangle  \right ) \Big],
 \\ \label{eq:vectoreq}
 Z_1= & \, 16\pi G  \langle \delta T_{0a}\rangle \hat {\bm \epsilon_a},
 \\ \label{eq:tensoreq}
 -\Box Z_2 
 = &\, \textstyle 16\pi G e^{-2 H t} \left [\langle \delta T_{ab}\rangle{-}\frac{1}{2} \langle \delta T_{cc}\rangle  \delta_{ab} \right ]\hat {\bm \epsilon}_a \otimes \hat {\bm \epsilon}_b,
\end{align}
\end{subequations}
where
\begin{equation}
\label{eq:box}
\Box \equiv g^{\mu \nu} \nabla_{\mu} \nabla_\nu = -[\partial_t^2 + 3 H \partial_t + q^2 e^{-2 H t}],
\end{equation}
is the scalar wave operator in de Sitter space.  The appropriate combinations of the linear semiclassical equations (\ref{eq:lineareinstein}) needed to derive Eqs.~(\ref{eq:boundaryinvariants}) can be read off from the coefficients of $\langle \delta T_{\mu \nu} \rangle$ appearing on the r.h.s. of these equations.

Naively Eqs.~(\ref{eq:Z0eq}) and (\ref{eq:vectoreq}) suggest $Z_0$ and $Z_1$ satisfy non-dynamical equations of motion.  Indeed, in four dimensional GR the only dynamical degrees of freedom are tensor modes.  However, SYM has its own dynamical excitations which can source the scalar and vector modes, rendering them dynamical.  In fact, we shall see below in Sec.~\ref{eq:5dpert} that the  $Z_s$ satisfy the same equations of motion for all helicities $s$, meaning all the $Z_s$ are dynamical in semiclassical gravity.

\section{Holographic calculation of $\langle T_{\mu \nu} \rangle $ and the semiclassical equations of motion}
\label{eq:holographic}

According to holographic duality \cite{Maldacena:1997re}, strongly coupled large $N$ SYM is equivalent to classical gravity in asymptotically AdS$_5$ spacetime.  
The 5D geometry is governed by the vacuum Einstein equations with cosmological constant $\Lambda_5 =-6/L^2$, with $L$ the AdS curvature radius.  The 5D geometry contains a 4D boundary, which is where the dual field theory lives.  The near-boundary asymptotics of the 5D metric encodes the SYM stress tensor $\langle T_{\mu \nu} \rangle$ \cite{Gubser:1998bc,Witten:1998qj,deHaro:2000vlm}.

\subsection{Near boundary asymptotics and $\langle T_{\mu \nu} \rangle $}

In what follows we will briefly review the analysis in Ref.~\cite{deHaro:2000vlm} for the computation of $\langle T_{\mu \nu} \rangle$. Following Ref.~\cite{deHaro:2000vlm}, we employ Fefferman-Graham coordinates where the 5D metric takes the form
\begin{equation}
\label{eq:FG}
ds^2 = L^2 \left [r^2 \g_{\mu \nu}(r,x^\alpha) dx^\mu dx^\nu + \frac{dr^2}{r^2} \right ],
\end{equation}
where $r$ is the AdS radial coordinate, with the boundary of the spacetime located at $r = \infty$. The coordinates $x^{\mu} = \{t,\bm x\}$ parameterize the boundary directions. The 5D Einstein equations imply $\frak g_{\mu \nu}(r,x^\alpha)$ has the near-boundary expansion
\begin{equation}
\label{eq:FGexpansion}
\g_{\alpha \beta} = \g_{\alpha \beta}^{(0)} + \g_{\alpha \beta}^{(2)}r^{-2} 
+ \g_{\alpha \beta}^{(4)}r^{-4}+ \h_{\alpha \beta}^{(4)} r^{-4} \log \left ( \textstyle  \frac{r}{\mu} \right) + O(r^{-5}).
\end{equation}
The expansion coefficients $\g_{\alpha \beta}^{(0)}$ and 
$\g_{\alpha \beta}^{(4)}$ are effectively constants of integration, with all other coefficients determined by them.
The scale $\mu$ appearing in the logarithm is an arbitrary scale related to renormalization in the dual SYM description.  Note that rescaling $\mu$ by a constant is equivalent to adding to $\g_{\alpha \beta}^{(4)}$ a term proportional to $\h_{\alpha \beta}^{(4)}$.  We discuss this further below.  

As a boundary condition 
we set
\begin{equation}
\g_{\mu \nu}^{(0)} = g_{\mu \nu},
\end{equation}
so the boundary metric is that where SYM lives. The 5D
Einstein equations then imply the coefficients $\g_{\mu\nu}^{(2)}$ and $\h_{\mu \nu}^{(4)}$ are determined by
curvature tensors of $g_{\mu \nu}$ via
\begin{align}
\g_{\mu \nu}^{(2)} &= \textstyle -\frac{1}{2} \left [R_{\mu \nu}- \frac{1}{6} R g_{\mu \nu} \right],
\\ \nonumber
\h_{\mu \nu}^{(4)} &= \textstyle \frac{1}{4} R_{\mu \alpha \nu \beta} R^{\alpha \beta} + \frac{1}{24} \del_\mu \del_\nu R - \frac{1}{8} \del^2 R_{\mu \nu} + \frac{1}{12} R R_{\mu \nu}
\\ 
&\textstyle +\frac{1}{16} \left ( R_{\alpha \beta} R^{\alpha \beta} + \frac{1}{3} \del^2 R - \frac{1}{3} R^2 \right ) g_{\mu \nu}, 
\end{align}
where $\del$ is the covariant derivative under the metric $g_{\mu \nu}$.  
Additionally, the 5D Einstein equations constrain the divergence and trace of $\g_{\mu \nu}^{(4)}$ to be
\begin{subequations}
	\label{eq:constraintsong4}
	\begin{align}
	\label{eq:cons}
	\textstyle \nabla^\mu \g_{\mu \nu}^{(4)} = &\, \nabla^\mu \big \{ \textstyle  \frac{1}{4} [ R^{\alpha \beta} R_{\alpha \beta} 
	- \frac{2}{3} R^2] g_{\mu \nu} \\ \nonumber
	&\, \textstyle - \frac{1}{8} [ R_{\mu \sigma} R^{\sigma}_{\ \nu} - \frac{1}{36} R^2 g_{\mu \nu}] \big\},
	\\
	\label{eq:g4trce}
	g^{\mu \nu} \g_{\mu \nu}^{(4)} = &\, \textstyle \frac{1}{16} \left (R^{\mu \nu}R_{\mu \nu} - \frac{2}{9} R^2 \right ).
	\end{align}
\end{subequations}
Note 
\begin{align}
\label{eq:hcons}
&g^{\mu \nu } \h_{\mu \nu}^{(4)} = 0, & \nabla^\mu \h_{\mu \nu}^{(4)} = 0,&
\end{align}
so any additive shift  to $\g_{\mu \nu}^{(4)}$ by $\h_{\mu \nu}^{(4)}$ does not affect Eqs.~(\ref{eq:constraintsong4}).   Hence, Eqs.~(\ref{eq:constraintsong4}) remain unchanged when the arbitrary renormalization scale $\mu$ is varied.
The remaining degrees of freedom in $\g_{\mu \nu}^{(4)}$ can only be determined by solving
Einstein's equations away from $r = \infty$.

The SYM stress tensor is given by the variation derivative of the 
renomalized 5D gravitational action $S_{\rm 5D}$ \cite{deHaro:2000vlm},
\begin{equation}
\label{eq:stressdef}
\langle T^{\mu \nu} \rangle =\frac{2}{\sqrt{-g}} \frac{\delta S_{\rm 5D}}{\delta g_{\mu \nu}}.
\end{equation}
Upon taking the variational derivative 
one finds \cite{deHaro:2000vlm}
\begin{align}
\nonumber
\langle T_{\mu \nu} \rangle =\ &  \frac{N^2}{2 \pi^2} \bigg [ \g_{\mu \nu}^{(4)} + \frac{1}{32} \left (R^{\alpha \beta} R_{\alpha \beta} - \frac{5}{9} R^2 \right ) g_{\mu \nu} \\  \label{eq:stress}
-\ & \frac{1}{8} R_{\mu \alpha}R^{\alpha \nu} + \frac{1}{16} R R_{\mu \nu} 
+  \alpha  \, \h_{\mu \nu}^{(4)} \bigg ].
\end{align}
Note that Eqs.~(\ref{eq:cons}) and (\ref{eq:stress}) imply the stress tensor satisfies the correct 
Weyl anomaly (\ref{eq:trace}) and is covariantly conserved,
\begin{equation}
\nabla_{\mu} \langle T^{\mu \nu} \rangle = 0.
\end{equation}

The last term in Eq.~(\ref{eq:stress}), $\alpha \, \h_{\mu \nu}^{(4)}$, arises from a counter term present in $S_{\rm 5D}$, which is necessary to cancel divergences near the boundary.  The salient counter term is proportional to the integral of the boundary Weyl anomaly (\ref{eq:trace}), meaning $S_{\rm 5D}$ contains a term proportional to
\begin{equation}
\label{eq:counterterm}
\int d^4 x \sqrt{-g}\left [ R^{\mu \nu} R_{\mu \nu} - \frac{1}{3} R^2 \right].
\end{equation} 
It turns out that up to an overall constant, $\h_{\mu \nu}^{(4)}$ is the contribution of the stress tensor from this term.  Correspondingly, $\alpha$ is the renormalized coupling  associated with the higher derivative correction (\ref{eq:counterterm}) to the 4D boundary gravitational action.

The coupling $\alpha$ in fact must be a function of the renormalization scale $\mu$.  Why?  Recall from Eq.~(\ref{eq:FGexpansion}) that 
rescaling $\mu$ by a constant results in an additive shift of $\h_{\mu \nu}^{(4)}$ to $\g_{\mu \nu}^{(4)}$.  It follows that  $\alpha$ and $\mu$ always appear in $\langle T_{\mu \nu} \rangle$ in the combination  
\begin{equation}
\label{eq:mudep}
{\textstyle \frac{N^2}{2 \pi^2}}\left  [\alpha + \log \mu \right ] \h_{\mu \nu}^{(4)}
\end{equation}
Since $\mu$ is arbitrary and the stress tensor is physical, this contribution to $\langle T_{\mu \nu} \rangle$ must be independent of $\mu$.  This requires $\alpha$ to run with $\mu$,
\begin{equation}
\alpha(\mu) = - \log \mu + \rm const.
\end{equation}
The energy scale
\begin{equation}
\label{eq:Edef}
\E \equiv \mu e^{\alpha},
\end{equation}
is thus a renormalization group invariant, meaning $\frac{dE}{d\mu} = 0$.  We emphasize that $\E$ is a physical energy scale in the quantum field theory, and that $\langle T_{\mu \nu} \rangle$ depends on $\E$ provided $\frak h_{\mu \nu}^{(4)}$ doesn't vanish.
In what follows we choose
\begin{equation}
\mu = \E.
\end{equation}
From Eq.~(\ref{eq:Edef}), it follows that $\alpha$ vanishes in this scheme.

\subsection{Vacuum state geometry}

What is the 5D geometry dual to the SYM vacuum state in de Sitter space?  This question can be answered by noting 
de Sitter space is conformally equivalent to Minkowski space.  The 5D geometry dual to the SYM vacuum in Minkowski space is AdS$_5$, whose metric reads
\begin{equation}
ds^2 = L^2 \tilde r^2 [-d \tilde t^2 + d \bm x^2] + \frac{L^2 d \tilde r^2}{\tilde r^2}.
\end{equation}
Consider then the coordinate transformation,
\begin{align}
\label{eq:coordtrans}
&\tilde t = \frac{e^{-H t} (H^2 + 4 r^2)}{H ( H^2 - 4 r^2)},&&
\tilde r = \frac{e^{H t} (4 r^2 - H^2)}{4 r}.&
\end{align}
On the boundary, this coordinate transformation is tantamount to a conformal transformation, mapping the Minkowski metric onto the de Sitter metric (\ref{eq:perturbeddS}) (with $h_{\mu\nu} = 0$).  In other words, on the boundary the transformation (\ref{eq:coordtrans}) maps the Minkowski vacuum onto the de Sitter vacuum.  Using the coordinate transformation (\ref{eq:coordtrans}), the 5D geometry dual to the de Sitter vacuum therefore reads \cite{Marolf:2010tg},
\begin{equation}
\label{eq:dsads}
ds^2 =  L^2 \left \{ \textstyle \frac{(4 r^2 - H^2)^2}{16 r^2} \left[ -dt^2 + e^{2 H t} d \bm x^2 \right ] + \frac{ dr^2}{r^2} \right \}.
\end{equation}
Note that the Poincare horizon  lies at $r = H/2$.

Comparing (\ref{eq:dsads}) with (\ref{eq:FGexpansion}), we see,
\begin{equation}
\g_{\mu \nu}^{(4)} = \frac{H^4}{16} g_{\mu \nu}.
\end{equation}
Using Eq.~(\ref{eq:stress}) we find,
\begin{equation}
\label{eq:desitterstress}
\langle T_{\mu \nu} \rangle = -\frac{3 N^2 H^4 }{32 \pi^2} g_{\mu \nu}.
\end{equation}
Hence, the vacuum stress is equivalent to a cosmological constant.
This reflects the fact that the vacuum of SYM is invariant under the symmetries of de Sitter space. Note that the vacuum stress is independent of the scale $\E$, which is a consequence of the fact that $\frak h_{\mu \nu}^{(4)}$ vanishes in de Sitter space.

\subsection{Bulk Metric perturbations and boundary semiclassical equations of motion}
\label{eq:5dpert}

Via Eq.~(\ref{eq:stress}), perturbations to the 5D metric (\ref{eq:dsads}) 
will induce perturbations $\langle \delta T_{\mu \nu} \rangle$ to the SYM 
stress tensor.
In Fefferman-Graham coordinates the perturbed 5D metric may be written as,
\begin{align}
\label{eq:dsadsperturbed}
\nonumber
ds^2 =   L^2  \bigg \{  \frac{(4 r^2 - H^2)^2}{16 r^2} & \bigg [ -(1 - \H_{00})dt^2 + 2 \H_{0i} dt dx^i \\
+ \ e^{2 H t} (&\delta_{ij} + \H_{ij}) dx^i dx^j  \bigg ] + \frac{ dr^2}{r^2}   \bigg \}.
\end{align}
where $\H_{\mu \nu}$ is the metric perturbation.  The linearized 5D Einstein equations must then be solved for $\H_{\mu\nu}$ subject to the boundary condition,
\begin{equation}
\label{eq:bc}
\lim_{r \to \infty} \H_{\mu \nu} = h_{\mu \nu},
\end{equation}
where $h_{\mu \nu}$ is the metric perturbation on the boundary.  A further 
boundary condition is need at the Poincare horizon.  Causality dictates that
no classical radiation can be emitted from $r = H/2$, meaning all gravitational waves near $r = H/2$ must be infalling only \cite{Son:2002sd}.

As done with the 4D Einstein equations in Sec~\ref{sec:bdinvariants}, instead of working directly with $\H_{\mu \nu}$ we choose
to work with gauge invariant combinations of $\H_{\mu \nu}$.
The construction of 5D invariants is essentially identical to that 
in 4D.  The salient gauge transformations are just the 5D generalization of Eqs.~(\ref{eq:difftrans0}) and (\ref{eq:difftrans}).  To construct 
gauge invariants we work with Fourier mode amplitudes $\H_{\mu \nu}(t,r,\bm q)$ and decompose $\H_{\mu \nu}(t,r,\bm q)$ in terms of longitudinal and transverse modes via Eq.~(\ref{eq:polarizationframe}).
In fact, it turns out the expressions (\ref{eq:gaugeinvariants}) for the boundary gauge invariants $Z_s$ are also diffeomorphism invariant in 5D with the replacement $h_{\mu \nu} \to \H_{\mu \nu}$.  
We therefore define the 5D helicity 0, 1 and 2 invariants 
\begin{subequations}
\begin{align}
\mathcal Z_0 = &\, \textstyle 2 q^2 e^{-2 H t} \H_{00} + 4 i q \, e^{-2 H t}[ \dot \H_{0q}- H \H_{0q}] 
\\ \nonumber
+ & \,\ddot \H_{aa}- 2 \ddot \H_{qq}+ H \dot \H_{aa} - 2 H \dot \H_{qq}+ e^{-2 H t} q^2 \H_{aa},
\\
\mathcal Z_1 = &\, [i q \dot \H_{aq}+ q^2 e^{-2 H t} \H_{0a}] \hat \epsilon_a ,
\\
\mathcal Z_2  = &\, [ \textstyle \H_{ab} - \frac{1}{2} \H_{cc} \delta_{ab} ] \hat \epsilon_a \otimes \hat \epsilon_b.
\end{align}
\end{subequations}

It is straightforward but tedious to show that the linearized 5D Einstein
equations imply $\mathcal Z_s$ satisfy the same wave equation for all helicities $s$,
\begin{align}
\label{eq:bulkevo}
\Box_{5} \mathcal Z_s  = 0,
\end{align}
where $\Box_{5}$ is the 5D scalar wave operator in the unperturbed geometry (\ref{eq:dsads}).
Explicitly,
\begin{align}
\label{eq:5Dwaveoperator}
\Box_{5} = \textstyle  \frac{16 r^2}{(H^2-4r^2)^2} \left \{ \Box  + \frac{r^3}{16 (H^2 - 4 r^2)^2 }\partial_r \frac{(H^2 - 4 r^2)^4}{r^3} \partial_r \right\},
\end{align}
where  again $\Box$ is the 4D scalar wave operator in de Sitter space defined in Eq.~(\ref{eq:box}).
The Feffermann-Graham expansion (\ref{eq:FGexpansion}) requires $\mathcal Z_s$ to have the near-boundary expansion
\begin{equation}
\label{eq:Zexp}
\mathcal Z_s = Z_s + \dots + \frac{1}{r^4} \mathcal Z_s^{(4)} + \dots,
\end{equation}
with all other terms in the expansion determined by the boundary value $Z_s$ and $\mathcal Z_s^{(4)}$.  The coefficient $\mathcal Z_s^{(4)}$ determines the sources on the r.h.s. of the boundary semiclassical equations of motion (\ref{eq:boundaryinvariants}).  In fact, Eqs.~(\ref{eq:boundaryinvariants}),
(\ref{eq:FGexpansion}), (\ref{eq:constraintsong4}) and (\ref{eq:stress}) imply each boundary gauge invariant $Z_s$ satisfies,
\begin{equation}
\label{eq:boundaryevo}
-\Box Z_s = \K \mathcal Z^{(4)}_s,
\end{equation}
where the constant $\K$ is given by,
\begin{equation}
\K \equiv \frac{8 G N^2}{\pi- G N^2 H^2}.
\end{equation}
The coupled scalar wave equations (\ref{eq:boundaryevo}) and (\ref{eq:bulkevo}) constitute the 
semiclassical equations of motion in SYM and are identical for all helicities $s$.

\section{Mode solutions to the semiclassical equations of motion}
\label{eq:modesolutions}

To solve the bulk equation of motion (\ref{eq:bulkevo}) we employ separation of variables, writing
\begin{equation}
\label{eq:seperation}
\mathcal Z_s(t,r) = Z_s(t) \mathcal F(r).
\end{equation}
The near-boundary expansion (\ref{eq:Zexp}) then requires the boundary condition
\begin{equation}
\label{eq:Fbc1}
\lim_{r\to \infty} \mathcal F(r) = 1.
\end{equation}
Denoting the separation constant as $H^2 (\nu^2 \textstyle - \frac{9}{4} )$ with $\nu$ a free parameter, the separated equations of motion read
\begin{equation}
\label{eq:bdevo2}
-\Box Z_s =  H^2 (\nu^2 \textstyle - \frac{9}{4} ) Z_s,
\end{equation}
and
\begin{align}
\label{eq:bulkev2o}
\left [ -H^2 (\nu^2 \textstyle {-} \frac{9}{4} )  + \frac{r^3}{16 (H^2 - 4 r^2)^2} \partial_r \frac{(H^2 - 4 r^2)^4}{r^3} \partial_r \right] \mathcal F = 0.
\end{align} 
Note the separated equations of motion  are invariant under $\nu \to -\nu$.
Correspondingly, for any fixed $\nu$ there are two sets of solutions related to each other by switching $\nu \to -\nu$.  Without loss of generality we focus on the $+\nu$ solution. 
 
Eq.~(\ref{eq:bdevo2}) is solved by,
\begin{equation}
Z_s = e^{-3 H t/2}  J_{\nu}( e^{-H t} q/H),
\end{equation}
where $J_\nu$ is a Bessel function.  At small $z$ we have $J_\nu(z) \sim z^{\nu}$.  
It follows that $Z_s$ remains bounded as $t \to \infty$ provided,
\begin{equation}
\label{eq:modestablity}
{\rm Re} \, \nu \geq -\frac{3}{2}.
\end{equation}
In what follows we shall employ the semiclassical equations of motion to compute the allowed values of $\nu$ and then use Eq.~(\ref{eq:modestablity})
as a criterion for mode stability.

The mode equation (\ref{eq:bulkev2o}) must be solved subject to the boundary condition (\ref{eq:Fbc1}) as well as the boundary condition that no radiation is emitted from the horizon at $r = \frac{H}{2}$.  
With these boundary conditions the solution to Eq.~(\ref{eq:bulkev2o}) reads,
\begin{equation}
\label{eq:sol}
\mathcal F =  \textstyle A  \left (\frac{r^2}{H^2}{-} \frac{1}{4} \right)^{-\nu - 3/2} {_2} F_1(-\frac{3}{2} {-} \nu,\frac{1}{2} {-} \nu,1-2 \nu,1 {-} \frac{4 r^2}{H^2}),
\end{equation}
where ${_2} F_1$ is a hypergeometric function and,
\begin{equation}
A \equiv \frac{2^{-2 \nu -3} \Gamma(\frac{1}{2} - \nu)\Gamma(\frac{5}{2} - \nu)}{\Gamma(1 - 2 \nu)},
\end{equation}
with $\Gamma(z)$ the gamma function.

Using (\ref{eq:seperation}) and expanding (\ref{eq:sol}) about $r = \infty$, we find
\begin{align}
\label{eq:Z4sol}
\mathcal Z_s^{(4)}  = \ & -\frac{ H^4 (\nu^2 - \frac{9}{4})Z_s }{256}  \bigg \{ 16
\\  \nonumber
+ \, & (1 - 4\nu^2)   \left [ 3 -4\mathscr H \left ( -{\textstyle  \frac{1}{2} - \nu}  \right ) +4 \log {\textstyle \frac{2 \E}{H}} \right ] \bigg \},
\end{align}
where 
\begin{equation}
\mathscr H(z) \equiv \frac{\Gamma'(z+1)}{\Gamma(z+1)} + \gamma_E,
\end{equation}
is the harmonic number function and $\gamma_E$ is the Euler-Mascheroni constant. Note the appearance of the energy scale $\E$, which in our renormalization scheme
is the same as the scale $\mu$ appearing in the Fefferman-Graham  expansion (\ref{eq:FGexpansion}).
Substituting (\ref{eq:Z4sol}) into the semiclassical equation of motion (\ref{eq:boundaryevo}) and using (\ref{eq:bdevo2}), we conclude the allowed values of $\nu$ must satisfy 
\begin{equation}
\label{eq:allowednu}
\textstyle(\nu^2 {-} \frac{9}{4}) Q(\nu) = 0,
\end{equation}
where 
\begin{align}
\nonumber
Q(\nu) \equiv &\, 256 + H^2 \K \Big \{ 16
\\ \label{eq:Qeq}
+ & \,  (1 - 4 \nu^2)   \left [ 3 - 4 \mathscr H \left ( -{\textstyle  \frac{1}{2} {-} \nu}  \right ) +4  \log {\textstyle \frac{2 \E}{H}} \right ] \Big \}.
\end{align}

There are infinitely many values of $\nu$ which satisfy Eq.~(\ref{eq:allowednu}).
We are interested in values which violate the bound (\ref{eq:modestablity}) and hence correspond to unstable modes. These values of $\nu$ must satisfy  
\begin{equation}
\label{eq:Qeq2}
	Q(\nu) = 0.
\end{equation}
  
We solve Eq.~(\ref{eq:Qeq2}) in the limit where both $G N^2 H^2 \ll 1$  and $G N^2 \E^2 \ll 1$.  This means $\mathcal K \approx 8 G N^2/\pi$.
Correspondingly, in the $G N^2 \to 0$ limit the quantity in the braces in Eq.~(\ref{eq:Qeq}) must diverge.  There are two way this can happen.  First, 
$\nu$ can be near the poles of $\mathscr H \left ( -{\textstyle  \frac{1}{2} - \nu}  \right )$.  However, these lie at positive half-integer values of $\nu$ and correspond to stable modes. Second, $\nu$ itself can diverge.
For large $|\nu|$ (and away from the positive real ${\rm Re }\, \nu$ axis) we may approximate
\begin{equation}
\label{eq:harmonicexpansion}
\mathscr H(\textstyle -\frac{1}{2}-\nu) = \log(-\nu)  + O(\nu^0).
\end{equation}
In the large $\nu$ limit Eq.~(\ref{eq:Qeq2}) then becomes
\begin{equation}
2 \pi - G N^2 H^2 \nu^2  \log \left (-\frac{2 \E}{H \nu} \right )  = 0.
\label{eq:Qeq3}
\end{equation}
Up to corrections suppressed by inverse powers of $\log \left (\frac{1}{G N^2 \E^2} \right )$, Eq.~(\ref{eq:Qeq3}) is solved by 
\begin{equation}
\label{eq:renu}
\nu = -\frac{ H_{\rm max}}{H} \sqrt{ \frac{1}{\log \left (\frac{1}{G N^2 \E^2} \right ) }}
	 \left [ i +\frac{\pi}{2\textstyle \log \left (\frac{1}{G N^2 \E^2}\right )} \right ],
\end{equation}
where the maximum Hubble constant $H_{\rm max}$ was defined in Eq.~(\ref{eq:Hmax}).

Eq.~(\ref{eq:renu}) and the stability condition (\ref{eq:modestablity}) imply de Sitter space with $H> H_c$ is mode stable whereas de Sitter space with $H< H_c$ is unstable.  The critical Hubble constant $H_c$ reads
\begin{equation}
H_c =   \frac{  \pi H_{\rm max} }{3} \left [\textstyle\frac{1}{  \log \left (\textstyle  \frac{1}{G N^2 \E^2} \right )} \right ]^{3/2}.
\end{equation}
Note that with $G N^2 \E^2 \ll 1$, $H_c$ is parameterically smaller than $H_{\rm max}$.

In the  $H \to 0$ limit, where de Sitter space becomes Minkowski space, the associated e-folding time $T$ of unstable modes is
\begin{equation}
\label{eq:timescaling}
T = \frac{2}{\pi H_{\rm max}} \left [\log  \frac{1}{G N^2 \E^2} \right ]^{3/2}.
\end{equation}
Note this time scale is parametrically larger than the Planck time.

\section{Discussion}

\label{sec:dis}

\begin{figure*}[ht]
	\includegraphics[trim= 0 0 0 0 ,clip,scale=0.6]{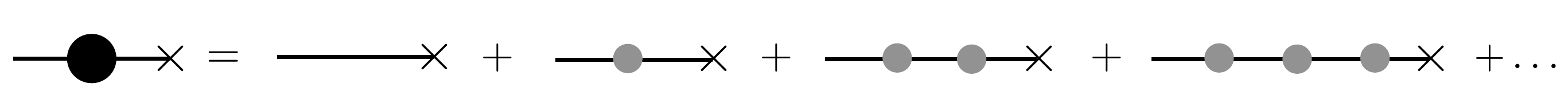}
	\caption{The diagramatic representation of the 
		perturbative expansion (\ref{eq:pertexp}), with $h_{\mu \nu}^{(n)}$ satisfying Eq.~(\ref{eq:perteq}). Each line segment represents the retarded Greens function associated with the operator $\Delta^{\mu \nu}_{\ \ \alpha \beta}$ and each light gray circle represent an insertion of $\Pi_{\mu \nu \alpha \beta}$, with all internal coordinates integrated over.  Likewise, the  $\times$ symbol represents classical stress tensor $\tau_{\mu \nu}$, with an integration over its coordinates as well.   The diagramatic expansion is simply that of the \textit{dressed} Greens function, which is the retarded Greens function of the linear operator $\bm \Delta - 8 \pi G \epsilon \bm \Pi$, convoluted with the classical stress tensor $\tau_{\mu \nu}$.
	}
	\label{fig:digram}
\end{figure*}

In this paper we studied mode stability of de Sitter space in semiclassical gravity with strongly coupled SYM.  We found that there exists a critical value of the Hubble constant $H_c$ for which de Sitter space with $H < H_c$ is unstable and $H > H_c$ is mode stable.  The fact that the Minkowski space limit is unstable agrees with  previous analyses of semiclassical gravity in Minkowski space with free quantum fields \cite{Horowitz:1978fq,Horowitz:1980fj,Hartle:1981zt,Suen:1989bg,Suen:1988uf,Jordan:1987wd,RandjbarDaemi:1981wd,Matsui:2019tlf,Matsui:2019tah}.

Given the small cosmological constant and associated Hubble constant in the real universe, it seems catastrophic that de Sitter space with small $H$ is unstable to exponentially growing modes, including those with inhomogeneities and anisotropies.  It is reasonable to surmise that this indicates a problem with semiclassical gravity, at least as it has been formulated in this and other papers which find similar instabilities. 

One potential solution to this problem is to treat the quantum stress $\langle \delta T_{\mu \nu} \rangle$ perturbatively.
To this end, consider the linearized semiclassical Einstein equations (\ref{eq:lineareinstein}) with a classical stress tensor $\tau_{\mu \nu}$ added to the r.h.s.,
\begin{equation}
\label{eq:lineareinstein2}
\Delta^{\alpha \beta}_{\ \ \mu \nu} h_{\alpha \beta} = 8 \pi G \epsilon \langle \delta T_{\mu \nu} \rangle + 8 \pi G \tau_{\mu \nu}.
\end{equation}
Notice we have introduced a bookkeeping parameter $\epsilon$ in front of $\langle \delta T_{\mu \nu} \rangle.$   
One can then expand the metric perturbation in powers of $\epsilon$,
\begin{equation}
\label{eq:pertexp}
h_{\mu \nu} = \sum_{n = 0}^\infty \epsilon^n h_{\mu \nu}^{(n)},
\end{equation}
and solve the equations of motion (\ref{eq:lineareinstein2}) order by order in $\epsilon$.
$h_{\mu \nu}^{(0)}$ just satisfies the classical linearized Einstein equations
sourced by $\tau_{\mu \nu}$.  To obtain the equations of motion for $h_{\mu \nu}^{(n)}$, we first employ linear response, 
\begin{equation}
\langle \delta T_{\mu \nu}(x) \rangle = \int d^4 x' \sqrt{-g} \, \Pi_{\mu \nu \alpha \beta}(x|x') h^{\alpha \beta}(x'),
\end{equation}
where $\Pi_{\mu \nu \alpha \beta}(x|x')$ is the retarded stress-stress correlator.
A short exercise then shows that $h_{\mu \nu}^{(n)}$ satisfies 
\begin{equation}
\label{eq:perteq}
\Delta^{\alpha \beta}_{\ \ \mu \nu} h_{\alpha \beta}^{(n)}(x) = 8 \pi G \int d^4 x'\sqrt{-g} \, \Pi_{\mu \nu}^{\ \ \alpha \beta}(x|x') h^{(n-1)}_{ \alpha \beta}(x').
\end{equation}
That is, $h_{\mu \nu}^{(n)}$ satisfies the 
linearized Einstein equations sourced by a linear functional of $h_{\mu \nu}^{(n-1)}$.  
This means that if we truncate the expansion (\ref{eq:pertexp}) at any finite order $n$, there will be no instability present.

Is truncating the expansion (\ref{eq:pertexp}) reasonable?  The expansion (\ref{eq:pertexp}) is simply that of the \textit{dressed} retarded Greens function of the linear operator $\bm \Delta - 8 \pi G \epsilon \bm \Pi$ convoluted with the classical stress tensor $\tau_{\mu \nu}$.  This is depicted diagramatically in Fig.~\ref{fig:digram}.  Accordingly, $\Pi_{\mu \nu \alpha \beta}$ is the graviton self energy.%
\footnote{In a perturbative treatment of quantum gravity there are also contributions to the graviton self energy from graviton loops.  However, these are $1/N$ suppressed relative to contributions from matter fields.}  As is well established in quantum field theory, the dispersion relation of physical excitations is encoded in dressed propagators.  Truncating perturbative expansions of propagators at finite order often leads to incorrect dispersion relations,\footnote{For example, the dispersion relation of electromagnetic waves in media can dramatically deviate from those in vacuum.} including the masking of physical instabilities.
We comment on this further below within the context of massless QED.  Aside from having the desirable feature of ameliorating instabilities in semiclassical gravity, we do not see any reason why the series in (\ref{eq:pertexp}) should be truncated at finite order or equivalently, why $\langle \delta T_{\mu \nu} \rangle$ should be treated perturbatively in the semiclassical equations of motion.

It is noteworthy that similar instabilities can also occur in semiclassical electrodynamics.  For example, in the Appendix we compute $\langle J_\mu \rangle$ in SYM coupled to a charged massless $\mathcal N = 2$ hypermultiplet \cite{Karch:2002sh}.  Using the semiclassical Maxwell equations (\ref{eq:macroem}), we find that electromagnetic instabilities exist in the $H \to 0$ limit.  In other words, the semiclassical Maxwell equations are unstable in Minkowski space. Conversely,  the semiclassical Maxwell equations are mode stable in the $H \to \infty$ limit.  Similar instabilities also occur in QED.
Understanding the origin of and resolution to instabilities in semiclassical electrodyanamics may serve as 
as useful guide to resolving those in semiclassical gravity.  

Consider an arbitrarily small and long wavelength perturbation to the QED vacuum (in Minkowski space with no backreaction on the geometry).  The one loop dispersion relation of electromagnetic excitations can be computed from location of poles in gauge field correlators or alternatively, from the semiclassical Maxwell equations 
(\ref{eq:macroem}) with $\langle J^\mu \rangle$ computed via linear response,
\begin{equation}
\label{eq:linresponseem}
\langle J^\mu (x) \rangle =  \int d^4 x' K^{\mu \nu}(x{-}x') A_\nu,
\end{equation}
where $K^{\mu \nu}$ is the retarded current-current correlator.  Introducing a spacetime Fourier transform with four momentum $q^\mu$, 
current conservation and Lorentz invariance requires 
\begin{equation}
\label{eq:jj}
K^{\mu \nu}(q) = \left [q^2 \eta^{\mu\nu} - q^\mu q^\nu \right] K(q),
\end{equation}
where $\eta^{\mu \nu}$ is the Minkowski space metric and $K(q)$
is a scalar function which encodes vacuum polarization.  Substituting (\ref{eq:linresponseem}) and (\ref{eq:jj}) into the equation of motion (\ref{eq:macroem}), it is easy to see that the dispersion relation of gauge invariant modes (i.e. electromagnetic fields) must satisfy 
\begin{equation}
\label{eq:emdispersion}
q^2\left [ \frac{1}{e^2} -  K(q) \right] = 0.
\end{equation}
 
$K$ can be computed perturbatively in QED (see e.g. \cite{brown_1992}).
At momenta much smaller than the electron mass, $K \to 0$, which reflects the fact that weak long wavelength electromagnetic fields cannot produce $e^+e^-$ pairs and hence a current $\langle J^\mu \rangle$.  Hence the only solution to (\ref{eq:emdispersion}) is $q^2 = 0$, which is simply the dispersion relation of propagating electromagnetic waves.
However, in the limit where the electron mass vanishes, 
\begin{equation}
\label{eq:Piscaling}
K \sim  \textstyle \log \frac{q^2}{\mu^2},
\end{equation}
where $\mu$ is an arbitrary scale.  Substituting (\ref{eq:Piscaling}) into (\ref{eq:emdispersion}), it is easy to see that 
there are solutions to (\ref{eq:emdispersion}) with ${\rm Im} \, q^0 > 0$.
%
In other words, in the massless limit there are exponentially growing modes.
This behavior is qualitatively similar to that discussed above in semiclassical gravity:   de Sitter space is stable only if the Hubble constant is suitably large.  

The presence of unstable modes indicates that the \textit{perturbative} ground state in massless QED is unstable.  Indeed, it was shown long ago that non-perturbative effects from soft electromagnetic modes alter the structure of the vacuum, including breaking symmetries \cite{Fomin:1978rk,Roberts:1985ju,Kapec:2017tkm}. In short, the presence of exponentially growing modes merely reflects the \textit{wrong} choice of vacuum to perturb around.  Notice that had we treated $\langle J_\mu \rangle$ perturbatively, as we outlined above for gravity, and truncated the resulting expansion for $A_\mu$ at finite order, we would have \textit{incorrectly} concluded that that the perturbative ground state is stable.  This further bolsters the notion that $\langle \delta T_{\mu \nu} \rangle$ should not be treated perturbatively in semiclassical gravity.


It is tempting to guess that a similar mechanism can exist in gravity, with soft gravitational modes non-perturbatively altering the structure of the vacuum when $H < H_c$.    
Given the small value of $H$ observed in the real universe, it is of considerable interest to explore the influence of non-perturbative effects on states with $H < H_c$.  Are these states stabilized? Or are the instabilities merely softened in some form? If so, are there residual instabilities which drive $H \to 0$, thereby explaining the origin of the present day small value of $H$?  Indeed, similar scenarios have been proposed in Refs.~\cite{PhysRevD.31.710,Polyakov:2012uc}.   Likewise, given that it appears the universe has transition from a period of large $H$ (inflation) to the present epoch of small $H$, it would be interesting to explore the possibility of quantum phase transitions in the early universe at $H = H_c$.  We leave these interesting questions for future study.

\begin{acknowledgments}
This work was supported by the Black Hole Initiative at Harvard University, 
which is funded by grants from the John Templeton Foundation and the Gordon and Betty Moore Foundation. We thank Gary Horowitz, Kevin Nguyen, Achilleas Porfyriadis, and Laurence Yaffe for useful discussions. 
\end{acknowledgments}

\appendix*
\section{Electromagnetic mode stability in de Sitter space}
\label{sec:app}

In this Appendix we employ holographic duality to compute $\langle J_\mu \rangle$ and then study mode stability of Maxwell's equations (\ref{eq:macroem}).  Our electromagnetic analysis  closely mimics the gravitational analysis presented in the body of the paper. 

The matter theory we consider is SYM coupled to a massless $\mathcal N = 2$ hypermultiplet.  For the Lagrangian of this theory see Ref.~\cite{Chesler:2006gr}.  With the addition of the massless hypermultiplet, the theory enjoys a $U(1)$
symmetry and an associated conserved current.  The $U(1)$ symmetry can be weakly gauged, thereby coupling the current to electrodynamics.  The 5D field dual to the current is a gauge field $\mathcal A_M$, whose dynamics are governed by the 5D vacuum Maxwell equations \cite{Karch:2002sh}.
These equations must be solved subject to the boundary condition that 
\begin{equation}
\label{eq:embc}
\lim_{r \to \infty} \mathcal A_\mu = A_\mu,
\end{equation}
where $A_\mu$ is the 4D vector potential in the field theory.  The expectation value of the current then reads
\begin{equation}
\langle J^\mu \rangle = \frac{1}{\sqrt{-g}} \frac{\delta S_{5D}}{\delta A_\mu},
\end{equation}
where $S_{5D}$ is the renormalized 5D gauge field action.
In the radial gauge $\mathcal A_r = 0$, the vector potential has the near-boundary expansion
\begin{equation}
\label{eq:emexp}
\mathcal A_\mu = \mathcal A_\mu^{(0)} + \frac{1}{r^2} \mathcal A_\mu^{(2)} + \dots,
\end{equation}
with $\mathcal A_\mu^{(0)} = A_\mu$.  In terms of these expansion coefficients, the expectation value of the current reads,
\begin{equation}
\label{eq:boundarycurrent}
\langle J_\mu \rangle = \frac{N}{4 \pi^2}\left [2 \mathcal A_\mu^{(2)}- \frac{1}{2} \nabla_\nu F_\mu^{\ \nu} \right ],
\end{equation}
where as usual $F_{\mu \nu}$ is the field strength of $A_\mu$.

As with our gravitational analysis, instead of working with the vector potentials $\mathcal A_M$ and $A_\mu$, we shall instead work with gauge invariant quantities.  We introduce a Fourier transform in the 3D spatial directions with momentum $\bm q$, and decompose the vector potentials
in terms of components parallel to and traverse to $\bm q$, just as done in Eq.~(\ref{eq:polarizationframe}) for metric perturbations.
We define the helicity 0 and 1  4D and 5D invariants (i.e. longitudinal and transverse electric fields),
\begin{align}
 E_0 \equiv &\,  \partial_t  A_q - i q  A_0, &
 E_1 \equiv &\,  A_a \hat {\bm \epsilon}_a,&
\\
\mathcal E_0 \equiv & \, \partial_t \mathcal A_q - i q \mathcal A_0, &
\mathcal E_1 \equiv & \, \mathcal A_a \hat {\bm \epsilon}_a.&
\end{align}
In de Sitter space the 4D Maxwell equations (\ref{eq:macroem}) then yield the following equations of motion,
\begin{subequations}
	\label{eq:boundaryE}
\begin{align}
\left [\textstyle \partial_t^2  + H \partial_t  + q^2 e^{-2 H t} \right] E_0 = &\,  e^2 \left [i q \langle J_0 \rangle + \partial_t  \langle J_q \rangle  \right ],
\\
\left [\textstyle \partial_t^2  + H \partial_t  + q^2 e^{-2 H t} \right] E_1 =&\, -2  e^2 \langle J_a \rangle \hat {\bm \epsilon}_a,
\end{align}
\end{subequations}
Likewise, in the geometry (\ref{eq:dsads}), the 5D vacuum Maxwell equations imply 
\begin{align}
\label{eq:5D}
\left [\textstyle \partial_t^2  + H \partial_t  + q^2 e^{-2 H t} 
-\frac{r}{16} \partial_r \frac{(H^2 - 4 r^2)^2}{r} \partial_r \right ]\mathcal E_s = 0,
\end{align}
for all $s$.

The expansion (\ref{eq:emexp}) and boundary condition (\ref{eq:embc}) implies
that near the boundary,
\begin{equation}
\label{eq:Eexp}
\mathcal E_s = E_s + \frac{1}{r^2} \mathcal E_s^{(2)} + \dots.
\end{equation}
It is straightforward to show from the 4D Maxwell equations (\ref{eq:boundaryE}), the definition of the current (\ref{eq:boundarycurrent}), and the boundary expansion (\ref{eq:emexp}),
that all $E_s$ satisfy
\begin{align}
\label{eq:Eseqm}
\left [\textstyle \partial_t^2  + H \partial_t  + q^2 e^{-2 H t} \right] E_s = \mathcal K_{EM} \mathcal E_s^{(2)},
\end{align}
where
\begin{equation}
\mathcal K_{EM} \equiv \frac{4 e^2 N}{ 8 \pi^2 - e^2 N}.
\end{equation}
Eqs.~(\ref{eq:5D}) and (\ref{eq:Eseqm}) constitute the semiclassical Maxwell equations.  They are identical
in form to the semiclassical gravity equations (\ref{eq:boundaryevo}) and (\ref{eq:bulkevo}). 

The 5D equation of motion (\ref{eq:5D}) can be solved with separation of variables,
\begin{equation}
\mathcal E_s(t,r) = E_s(t) \mathcal F(r).
\end{equation}
Denoting the separation constant by $\textstyle H^2 \left (\nu^2 - \frac{1}{4}\right )$, where $\nu$ is a free parameter, the separated equations of motion read,
\begin{equation}
\label{eq:boundaryeigen}
\left [\textstyle \partial_t^2  + H \partial_t  + q^2 e^{-2 H t} \right] E_s = \textstyle H^2 (\nu^2 - \frac{1}{4}) E_s.
\end{equation}
and
\begin{align}
\label{eq:bulkeigen}
\left [\textstyle H^2 \left (\nu^2 - \frac{1}{4}\right ) 
-\frac{r}{16} \partial_r \frac{(H^2 - 4 r^2)^2}{r} \partial_r \right ]\mathcal F = 0.
\end{align}
Eq.~(\ref{eq:boundaryeigen}) is solved by $e^{-H t/2} J_\nu(e^{- Ht} q/H)$ where again $J_\nu$ is a Bessel function.  Having $E_s$ remain bounded as $t \to \infty$ then requires
\begin{equation}
\label{eq:nuboundem}
{\rm Re} \, \nu \geq -\frac{1}{2}.
\end{equation}

Eq.~(\ref{eq:bulkeigen}) must be solved subject to the boundary condition 
$\lim_{r \to \infty} \mathcal F(r) = 1$.  With the further boundary condition that there are no waves emitted from the horizon at $r = \frac{H}{2}$, the solution to (\ref{eq:bulkeigen}) reads,
\begin{equation}
\label{eq:emsol}
\mathcal F =  \textstyle A  \left (1-\frac{H^2}{4 r^2} \right)^{1/2 - \nu} {_2} F_1(-\frac{1}{2} {-} \nu,\frac{1}{2} {-} \nu,1-2 \nu,1 {-} \frac{4 r^2}{H^2}),
\end{equation}
where again ${_2} F_1$ is a hypergeometric function, and
\begin{equation}
A = \frac{\Gamma(1 - 2 \nu)}{\Gamma(\frac{1}{2} - \nu) \Gamma(\frac{3}{2} - \nu)}.
\end{equation}
Expanding Eq.~(\ref{eq:emsol}) about $r = \infty$, we find
to compute $\mathcal E_s^{(2)}$, and using (\ref{eq:boundaryeigen}) and (\ref{eq:Eseqm}),
we find that unstable modes must satisfy 
\begin{equation}
\label{eq:unstableemmodes}
4 + \textstyle \mathcal K_{EM} \left [1 - 2 \mathscr H \left ( -\frac{1}{2} -\nu \right ) + 2 \log \frac{2 \mu}{H} \right ] = 0.
\end{equation}
Note the appearance of the arbitrary renormalization scale $\mu$.  The $\mu$ dependence in (\ref{eq:unstableemmodes}) must be canceled by that of the running coupling $e(\mu)$. 

What are the allowed values of $\nu$ in Eq.~(\ref{eq:unstableemmodes})?  First, consider the limit $H \to 0$.  In this case the logarithm in (\ref{eq:unstableemmodes}) is large and positive, meaning $\mathscr H \left ( -\frac{1}{2} -\nu \right )$ must be large and positive.  This happens near the poles of $\mathscr H \left ( -\frac{1}{2} -\nu \right )$, which occur at positive half-integer values of $\nu$ and correspond to stable modes, or at large and negative $\nu$.
Using the expansion (\ref{eq:harmonicexpansion}),  we find that the associated $\nu$ are given by
\begin{equation}
\nu = - \frac{2 \mu}{H} \exp\left [\frac{1}{2} - \gamma_E + \frac{1}{2 \mathcal K_{EM}} \right ].
\end{equation}
This value of $\nu$ violates the bound (\ref{eq:nuboundem}), indicating an instability in the $H \to 0$ limit.

Conversely, consider the $H \to \infty$ limit.  In this case the logarithm in (\ref{eq:unstableemmodes}) is large and negative, meaning $\mathscr H \left ( -\frac{1}{2} -\nu \right )$ must be large and negative.  This only happens near the poles of $\mathscr H \left ( -\frac{1}{2} -\nu \right )$, which again are at positive half-integer values of $\nu$ and correspond to stable modes.  In other words, Maxwell's equations (\ref{eq:macroem}) are mode stable in the $H \to \infty$ limit.  Just like in our gravitational analysis, it follows there must exist a critical value of $H$ which separates the stable sector of the theory from the unstable sector. 

\bibliographystyle{utphys}
\bibliography{refs}%

\providecommand{\href}[2]{#2}\begingroup\raggedright\begin{thebibliography}{10}

\bibitem{kapusta_gale_2006}
J.~I. Kapusta and C.~Gale,
  \href{http://dx.doi.org/10.1017/CBO9780511535130}{{\em Finite-Temperature
  Field Theory: Principles and Applications}}.
\newblock Cambridge Monographs on Mathematical Physics. Cambridge University
  Press, 2~ed., 2006.

\bibitem{fetter2012quantum}
A.~Fetter and J.~Walecka, {\em Quantum Theory of Many-Particle Systems}.
\newblock Dover Books on Physics. Dover Publications, 2012.
\newblock \url{https://books.google.com/books?id=t5\_DAgAAQBAJ}.

\bibitem{birrell_davies_1982}
N.~D. Birrell and P.~C.~W. Davies,
  \href{http://dx.doi.org/10.1017/CBO9780511622632}{{\em Quantum Fields in
  Curved Space}}.
\newblock Cambridge Monographs on Mathematical Physics. Cambridge University
  Press, 1982.

\bibitem{friedrich1986}
H.~Friedrich, ``On the existence of $n$-geodesically complete or future
  complete solutions of einstein's field equations with smooth asymptotic
  structure,'' {\em Comm. Math. Phys.} {\bf 107} (1986) no.~4, 587--609.
  \url{https://projecteuclid.org:443/euclid.cmp/1104116232}.

\bibitem{FRIEDRICH1986101}
H.~Friedrich, ``Existence and structure of past asymptotically simple solutions
  of einstein's field equations with positive cosmological constant,''
  \href{http://dx.doi.org/https://doi.org/10.1016/0393-0440(86)90004-5}{{\em
  Journal of Geometry and Physics} {\bf 3} (1986) no.~1, 101 -- 117}.
  \url{http://www.sciencedirect.com/science/article/pii/0393044086900045}.

\bibitem{Anderson:2004ir}
M.~T. Anderson, ``{Existence and stability of even dimensional asymptotically
  de Sitter spaces},'' \href{http://dx.doi.org/10.1007/s00023-005-0224-x}{{\em
  Annales Henri Poincare} {\bf 6} (2005)  801--820},
\href{http://arxiv.org/abs/gr-qc/0408072}{{\tt arXiv:gr-qc/0408072 [gr-qc]}}.

\bibitem{Christodoulou:1993uv}
D.~Christodoulou and S.~Klainerman,
``{The Global nonlinear stability of the Minkowski space},''.

\bibitem{Lindblad:2004ue}
H.~Lindblad and I.~Rodnianski, ``{The Global stability of the Minkowski
  space-time in harmonic gauge},''
\href{http://arxiv.org/abs/math/0411109}{{\tt arXiv:math/0411109 [math-ap]}}.

\bibitem{Horowitz:1978fq}
G.~T. Horowitz and R.~M. Wald, ``{Dynamics of Einstein's Equation Modified by a
  Higher Order Derivative Term},''
\href{http://dx.doi.org/10.1103/PhysRevD.17.414}{{\em Phys. Rev.} {\bf D17}
  (1978)  414--416}.

\bibitem{Horowitz:1980fj}
G.~T. Horowitz, ``{SEMICLASSICAL RELATIVITY: THE WEAK FIELD LIMIT},''
\href{http://dx.doi.org/10.1103/PhysRevD.21.1445}{{\em Phys. Rev.} {\bf D21}
  (1980)  1445--1461}.

\bibitem{Hartle:1981zt}
J.~B. Hartle and G.~T. Horowitz, ``{Ground State Expectation Value of the
  Metric in the 1/$N$ or Semiclassical Approximation to Quantum Gravity},''
\href{http://dx.doi.org/10.1103/PhysRevD.24.257}{{\em Phys. Rev.} {\bf D24}
  (1981)  257--274}.

\bibitem{Suen:1989bg}
W.~M. Suen, ``{Minkowski Space-time Is Unstable in Semiclassical Gravity},''
\href{http://dx.doi.org/10.1103/PhysRevLett.62.2217}{{\em Phys. Rev. Lett.}
  {\bf 62} (1989)  2217--2220}.

\bibitem{Suen:1988uf}
W.-M. Suen, ``{The Stability of the Semiclassical Einstein Equation},''
\href{http://dx.doi.org/10.1103/PhysRevD.40.315}{{\em Phys. Rev.} {\bf D40}
  (1989)  315}.

\bibitem{Jordan:1987wd}
R.~D. Jordan, ``{Stability of Flat Space-time in Quantum Gravity},''
\href{http://dx.doi.org/10.1103/PhysRevD.36.3593}{{\em Phys. Rev.} {\bf D36}
  (1987)  3593--3603}.

\bibitem{RandjbarDaemi:1981wd}
S.~Randjbar-Daemi, ``{Stability of the Minkowski Vacuum in the Renormalized
  Semiclassical Theory of Gravity},''
\href{http://dx.doi.org/10.1088/0305-4470/14/7/001}{{\em J. Phys.} {\bf A14}
  (1981)  L229}.

\bibitem{Matsui:2018iez}
H.~Matsui, ``{Instability of De Sitter Spacetime induced by Quantum Conformal
  Anomaly},'' \href{http://dx.doi.org/10.1088/1475-7516/2019/01/003}{{\em JCAP}
  {\bf 1901} (2019) no.~01, 003},
\href{http://arxiv.org/abs/1806.10339}{{\tt arXiv:1806.10339 [hep-th]}}.

\bibitem{Matsui:2019tlf}
H.~Matsui and N.~Watamura, ``{Quantum Spacetime Instability and Breakdown of
  Semiclassical Gravity},''
\href{http://arxiv.org/abs/1910.02186}{{\tt arXiv:1910.02186 [gr-qc]}}.

\bibitem{Matsui:2019tah}
H.~Matsui, ``{Spacetime Instability and the Problems with Low Energy Quantum
  Gravity},''
\href{http://arxiv.org/abs/1901.08785}{{\tt arXiv:1901.08785 [hep-th]}}.

\bibitem{Anderson:2002fk}
P.~R. Anderson, C.~Molina-Paris, and E.~Mottola, ``{Linear response, validity
  of semiclassical gravity, and the stability of flat space},''
  \href{http://dx.doi.org/10.1103/PhysRevD.67.024026}{{\em Phys. Rev.} {\bf
  D67} (2003)  024026},
\href{http://arxiv.org/abs/gr-qc/0209075}{{\tt arXiv:gr-qc/0209075 [gr-qc]}}.

\bibitem{Polyakov:2012uc}
A.~M. Polyakov, ``{Infrared instability of the de Sitter space},''
\href{http://arxiv.org/abs/1209.4135}{{\tt arXiv:1209.4135 [hep-th]}}.

\bibitem{Maldacena:1997re}
J.~M. Maldacena, ``{The Large N limit of superconformal field theories and
  supergravity},'' \href{http://dx.doi.org/10.1023/A:1026654312961,
  10.4310/ATMP.1998.v2.n2.a1}{{\em Int. J. Theor. Phys.} {\bf 38} (1999)
  1113--1133}, \href{http://arxiv.org/abs/hep-th/9711200}{{\tt
  arXiv:hep-th/9711200 [hep-th]}}.
[Adv. Theor. Math. Phys.2,231(1998)].

\bibitem{Maldacena:2012xp}
J.~Maldacena and G.~L. Pimentel, ``{Entanglement entropy in de Sitter space},''
  \href{http://dx.doi.org/10.1007/JHEP02(2013)038}{{\em JHEP} {\bf 02} (2013)
  038},
\href{http://arxiv.org/abs/1210.7244}{{\tt arXiv:1210.7244 [hep-th]}}.

\bibitem{Fischler:2013fba}
W.~Fischler, S.~Kundu, and J.~F. Pedraza, ``{Entanglement and
  out-of-equilibrium dynamics in holographic models of de Sitter QFTs},''
  \href{http://dx.doi.org/10.1007/JHEP07(2014)021}{{\em JHEP} {\bf 07} (2014)
  021},
\href{http://arxiv.org/abs/1311.5519}{{\tt arXiv:1311.5519 [hep-th]}}.

\bibitem{Fischler:2014ama}
W.~Fischler, P.~H. Nguyen, J.~F. Pedraza, and W.~Tangarife, ``{Holographic
  Schwinger effect in de Sitter space},''
  \href{http://dx.doi.org/10.1103/PhysRevD.91.086015}{{\em Phys. Rev.} {\bf
  D91} (2015) no.~8, 086015},
\href{http://arxiv.org/abs/1411.1787}{{\tt arXiv:1411.1787 [hep-th]}}.

\bibitem{Fischler:2014tka}
W.~Fischler, P.~H. Nguyen, J.~F. Pedraza, and W.~Tangarife, ``{Fluctuation and
  dissipation in de Sitter space},''
  \href{http://dx.doi.org/10.1007/JHEP08(2014)028}{{\em JHEP} {\bf 08} (2014)
  028},
\href{http://arxiv.org/abs/1404.0347}{{\tt arXiv:1404.0347 [hep-th]}}.

\bibitem{Nguyen:2017ggc}
K.~Nguyen, ``{De Sitter-invariant States from Holography},''
  \href{http://dx.doi.org/10.1088/1361-6382/aae86b}{{\em Class. Quant. Grav.}
  {\bf 35} (2017) no.~22, 225006},
\href{http://arxiv.org/abs/1710.04675}{{\tt arXiv:1710.04675 [hep-th]}}.

\bibitem{Hawking:2000bb}
S.~W. Hawking, T.~Hertog, and H.~S. Reall, ``{Trace anomaly driven
  inflation},'' \href{http://dx.doi.org/10.1103/PhysRevD.63.083504}{{\em Phys.
  Rev.} {\bf D63} (2001)  083504},
\href{http://arxiv.org/abs/hep-th/0010232}{{\tt arXiv:hep-th/0010232
  [hep-th]}}.

\bibitem{CasalderreySolana:2011us}
J.~Casalderrey-Solana, H.~Liu, D.~Mateos, K.~Rajagopal, and U.~A. Wiedemann,
  ``{Gauge/String Duality, Hot QCD and Heavy Ion Collisions},''
\href{http://arxiv.org/abs/1101.0618}{{\tt arXiv:1101.0618 [hep-th]}}.

\bibitem{Hartnoll:2016apf}
S.~A. Hartnoll, A.~Lucas, and S.~Sachdev, ``{Holographic quantum matter},''
\href{http://arxiv.org/abs/1612.07324}{{\tt arXiv:1612.07324 [hep-th]}}.

\bibitem{Tomboulis:1977jk}
E.~Tomboulis, ``{1/N Expansion and Renormalization in Quantum Gravity},''
\href{http://dx.doi.org/10.1016/0370-2693(77)90678-5}{{\em Phys. Lett.} {\bf
  70B} (1977)  361--364}.

\bibitem{Kuo:1993if}
C.-I. Kuo and L.~H. Ford, ``{Semiclassical gravity theory and quantum
  fluctuations},'' \href{http://dx.doi.org/10.1103/PhysRevD.47.4510}{{\em Phys.
  Rev.} {\bf D47} (1993)  4510--4519},
\href{http://arxiv.org/abs/gr-qc/9304008}{{\tt arXiv:gr-qc/9304008 [gr-qc]}}.

\bibitem{Kovtun:2005ev}
P.~K. Kovtun and A.~O. Starinets, ``{Quasinormal modes and holography},''
  \href{http://dx.doi.org/10.1103/PhysRevD.72.086009}{{\em Phys. Rev.} {\bf
  D72} (2005)  086009},
\href{http://arxiv.org/abs/hep-th/0506184}{{\tt arXiv:hep-th/0506184
  [hep-th]}}.

\bibitem{Chesler:2007sv}
P.~M. Chesler and L.~G. Yaffe, ``{The Stress-energy tensor of a quark moving
  through a strongly-coupled N=4 supersymmetric Yang-Mills plasma: Comparing
  hydrodynamics and AdS/CFT},''
  \href{http://dx.doi.org/10.1103/PhysRevD.78.045013}{{\em Phys. Rev.} {\bf
  D78} (2008)  045013},
\href{http://arxiv.org/abs/0712.0050}{{\tt arXiv:0712.0050 [hep-th]}}.

\bibitem{Chesler:2007an}
P.~M. Chesler and L.~G. Yaffe, ``{The Wake of a quark moving through a
  strongly-coupled plasma},''
  \href{http://dx.doi.org/10.1103/PhysRevLett.99.152001}{{\em Phys. Rev. Lett.}
  {\bf 99} (2007)  152001},
\href{http://arxiv.org/abs/0706.0368}{{\tt arXiv:0706.0368 [hep-th]}}.

\bibitem{Hong:2011bd}
J.~Hong, D.~Teaney, and P.~M. Chesler, ``{The Wake of a Heavy Quark in
  Non-Abelian Plasmas : Comparing Kinetic Theory and the AdS/CFT
  Correspondence},'' \href{http://dx.doi.org/10.1103/PhysRevC.85.064903}{{\em
  Phys. Rev.} {\bf C85} (2012)  064903},
\href{http://arxiv.org/abs/1110.5292}{{\tt arXiv:1110.5292 [nucl-th]}}.

\bibitem{Duff:1993wm}
M.~J. Duff, ``{Twenty years of the Weyl anomaly},''
  \href{http://dx.doi.org/10.1088/0264-9381/11/6/004}{{\em Class. Quant. Grav.}
  {\bf 11} (1994)  1387--1404},
\href{http://arxiv.org/abs/hep-th/9308075}{{\tt arXiv:hep-th/9308075
  [hep-th]}}.

\bibitem{Henningson:1998ey}
M.~Henningson and K.~Skenderis, ``{Holography and the Weyl anomaly},''
  \href{http://dx.doi.org/10.1002/(SICI)1521-3978(20001)48:1/3<125::AID-PROP125>3.0.CO;2-B,
  10.1002/(SICI)1521-3978(20001)48:1/3<125::AID-PROP125>3.3.CO;2-2}{{\em
  Fortsch. Phys.} {\bf 48} (2000)  125--128},
\href{http://arxiv.org/abs/hep-th/9812032}{{\tt arXiv:hep-th/9812032
  [hep-th]}}.

\bibitem{Gubser:1998bc}
S.~S. Gubser, I.~R. Klebanov, and A.~M. Polyakov, ``{Gauge theory correlators
  from noncritical string theory},''
  \href{http://dx.doi.org/10.1016/S0370-2693(98)00377-3}{{\em Phys. Lett.} {\bf
  B428} (1998)  105--114},
\href{http://arxiv.org/abs/hep-th/9802109}{{\tt arXiv:hep-th/9802109
  [hep-th]}}.

\bibitem{Witten:1998qj}
E.~Witten, ``{Anti-de Sitter space and holography},''
  \href{http://dx.doi.org/10.4310/ATMP.1998.v2.n2.a2}{{\em Adv. Theor. Math.
  Phys.} {\bf 2} (1998)  253--291},
\href{http://arxiv.org/abs/hep-th/9802150}{{\tt arXiv:hep-th/9802150
  [hep-th]}}.

\bibitem{deHaro:2000vlm}
S.~de~Haro, S.~N. Solodukhin, and K.~Skenderis, ``{Holographic reconstruction
  of space-time and renormalization in the AdS / CFT correspondence},''
  \href{http://dx.doi.org/10.1007/s002200100381}{{\em Commun. Math. Phys.} {\bf
  217} (2001)  595--622},
\href{http://arxiv.org/abs/hep-th/0002230}{{\tt arXiv:hep-th/0002230
  [hep-th]}}.

\bibitem{Marolf:2010tg}
D.~Marolf, M.~Rangamani, and M.~Van~Raamsdonk, ``{Holographic models of de
  Sitter QFTs},'' \href{http://dx.doi.org/10.1088/0264-9381/28/10/105015}{{\em
  Class. Quant. Grav.} {\bf 28} (2011)  105015},
\href{http://arxiv.org/abs/1007.3996}{{\tt arXiv:1007.3996 [hep-th]}}.

\bibitem{Son:2002sd}
D.~T. Son and A.~O. Starinets, ``{Minkowski space correlators in AdS / CFT
  correspondence: Recipe and applications},''
  \href{http://dx.doi.org/10.1088/1126-6708/2002/09/042}{{\em JHEP} {\bf 09}
  (2002)  042},
\href{http://arxiv.org/abs/hep-th/0205051}{{\tt arXiv:hep-th/0205051
  [hep-th]}}.

\bibitem{Karch:2002sh}
A.~Karch and E.~Katz, ``{Adding flavor to AdS / CFT},''
  \href{http://dx.doi.org/10.1088/1126-6708/2002/06/043}{{\em JHEP} {\bf 06}
  (2002)  043},
\href{http://arxiv.org/abs/hep-th/0205236}{{\tt arXiv:hep-th/0205236
  [hep-th]}}.

\bibitem{brown_1992}
L.~S. Brown, \href{http://dx.doi.org/10.1017/CBO9780511622649}{{\em Quantum
  Field Theory}}.
\newblock Cambridge University Press, 1992.

\bibitem{Fomin:1978rk}
P.~I. Fomin, V.~P. Gusynin, and V.~A. Miransky, ``{Vacuum Instability of
  Massless Electrodynamics and the Gell-mann-low Eigenvalue Condition for the
  Bare Coupling Constant},''
\href{http://dx.doi.org/10.1016/0370-2693(78)90366-0}{{\em Phys. Lett.} {\bf
  78B} (1978)  136--139}.

\bibitem{Roberts:1985ju}
C.~D. Roberts and R.~T. Cahill, ``{Dynamically Selected Vacuum Field
  Configuration in Massless {QED}},''
\href{http://dx.doi.org/10.1103/PhysRevD.33.1755}{{\em Phys. Rev.} {\bf D33}
  (1986)  1755}.

\bibitem{Kapec:2017tkm}
D.~Kapec, M.~Perry, A.-M. Raclariu, and A.~Strominger, ``{Infrared Divergences
  in QED, Revisited},''
  \href{http://dx.doi.org/10.1103/PhysRevD.96.085002}{{\em Phys. Rev.} {\bf
  D96} (2017) no.~8, 085002},
\href{http://arxiv.org/abs/1705.04311}{{\tt arXiv:1705.04311 [hep-th]}}.

\bibitem{PhysRevD.31.710}
L.~H. Ford, \href{http://dx.doi.org/10.1103/PhysRevD.31.710}{``Quantum
  instability of de sitter spacetime,''{\em Phys. Rev. D} {\bf 31} (Feb, 1985)
  710--717}. \url{https://link.aps.org/doi/10.1103/PhysRevD.31.710}.

\bibitem{Chesler:2006gr}
P.~M. Chesler and A.~Vuorinen, ``{Heavy flavor diffusion in weakly coupled N=4
  super Yang-Mills theory},''
  \href{http://dx.doi.org/10.1088/1126-6708/2006/11/037}{{\em JHEP} {\bf 11}
  (2006)  037},
\href{http://arxiv.org/abs/hep-ph/0607148}{{\tt arXiv:hep-ph/0607148
  [hep-ph]}}.

\end{thebibliography}\endgroup

\end{document}